\documentclass[epj]{svjour}
\usepackage{graphicx}
\usepackage{amsmath}
\usepackage{cite}
\usepackage{amssymb}
\usepackage{threeparttable}
\begin{document}
\title{$\alpha$ + core structure described with an additional interaction in the nuclear matter saturation region}
\titlerunning{$\alpha$ + core structure described with an additional interaction...}
\author{M.~A.~Souza \thanks{E-mail: \texttt{marsouza@if.usp.br}}
\and H.~Miyake \thanks{E-mail: \texttt{miyake@if.usp.br}}%
%
}                     
%
%
\institute{Instituto de F\'{\i}sica, Universidade de S\~{a}o Paulo, Rua do Mat\~{a}o, 1371, CEP 05508-090,
 Cidade Universit\'{a}ria, \\ S\~{a}o Paulo - SP, Brazil}
\date{Received: date / Revised version: date}
%
\abstract{
In a phenomenological approach, the $\alpha$ + core structure is investigated in the $^{20}$Ne, $^{44}$Ti, $^{94}$Mo, $^{104}$Te, and $^{212}$Po nuclei through the local potential model using a double-folding nuclear potential with effective nucleon-nucleon interaction of \mbox{M3Y + $c_{\mathrm{sat}}\delta(s)$} type, where the term $c_{\mathrm{sat}}\delta(s)$ acts only between the saturation regions of $\alpha$-cluster and core. Properties such as energy levels, $\alpha$-widths, $B(E2)$ transition rates, and half-lives are calculated, and good level of agreement with experimental data is obtained in general. It is shown the inclusion of the term $c_{\mathrm{sat}}\delta(s)$ is determinant for a better description of the experimental energy levels compared to the ground state bands produced with the simple M3Y interaction. The properties calculated for $^{104}$Te reinforce the superallowed $\alpha$-decay feature for this nucleus, as indicated in previous studies.
} 
\maketitle
\section{Introduction}
\label{Sec:introduction}

The cluster structure of the atomic nucleus has become one of the main topics in nuclear physics. Its applications are seen in the studies on nuclear structure, spectroscopic properties, nuclear reactions, $\alpha$ or exotic nuclear decay, and astrophysical phenomena. The cluster models are developed with different configurations, ranging from the binary \mbox{cluster + core} structure to molecular structures based on $\alpha$-particles or $\alpha$-particles + nucleons. Cluster + core binary systems can be investigated by more direct approaches, such as the local potential model (LPM); on the other hand, the cluster structures can be analyzed by microscopic approaches such as the orthogonality condition model (OCM), the antisymmetrized molecular dynamics (AMD), the resonating group method (RGM), the quantum Monte Carlo method (QMC), and others, mainly in light nuclei \cite{FHK2018}. Different experiments have been carried out to corroborate the theoretical predictions, as shown in recent publications (e.g., \cite{TRS2021,WZM2022,BKF2021,DMC2021}).

In the last decades, there was an effort by different authors to describe the $\alpha$ + core structure through a systematics applicable to a set of nuclei, mainly nuclei with $\alpha$-clustering above double-shell closures (e.g., Refs.~\cite{BMP95,WPX2013,SMB2019,IMP2019,BR2021,SM2015,M2008,NR2011,SM2021}). In general, these works yield satisfactory results for energy spectra, electromagnetic transition rates, $\alpha$-widths, half-lives, and rms charge radii; however, for some nuclear potential forms, the description of energy levels is unsatisfactory, requiring the use of a variable intensity parameter (in most cases, dependent on the angular momentum $L$) to obtain energy bands closer to the experimental levels. This feature is observed in the energy bands produced by the double-folding nuclear potentials, mainly in the heavier nuclei. Double-folding nuclear potentials are known to be successful in reproducing $\alpha$-nucleus elastic scattering cross sections in a wide range of masses and energies, as shown in several works using the DDM3Y interaction \cite{KBL1984,O1995,MKF2013,M2017,AMA1996}; however, the double-folding potentials produce energy bands with strong rotational feature, which is incompatible with experimental spectra in many cases; in addition, for heavier nuclei, double-folding potentials can generate very compressed energy bands compared to the experimental levels, making it necessary to use the $L$-dependent intensity parameter.

\begin{figure*}
\begin{center}
\vspace{2mm}
\includegraphics[scale=0.78]{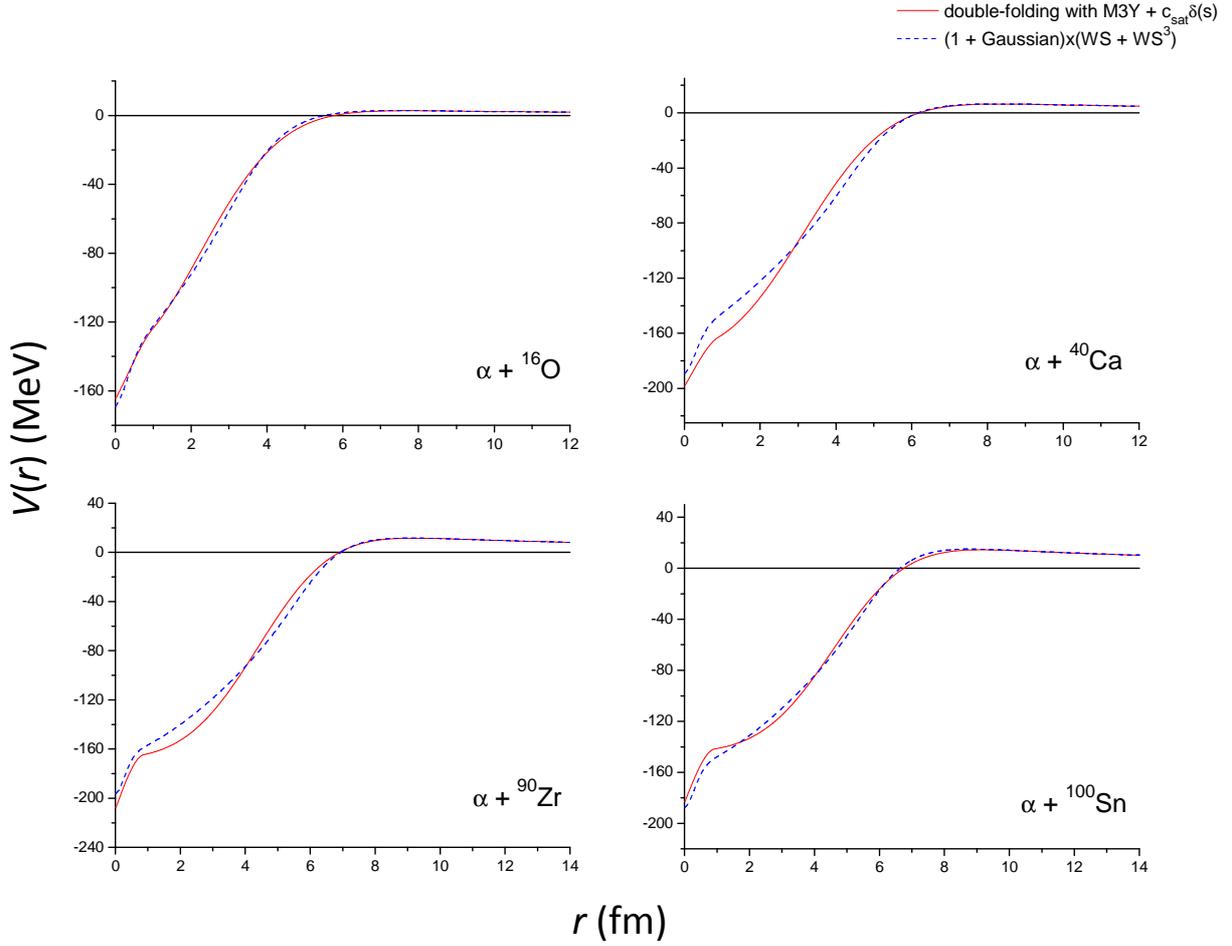}
\end{center}
\caption{Potential $V(r)$ applied to the $\alpha + ^{16}$O, $\alpha + ^{40}$Ca, $\alpha + ^{90}$Zr, and $\alpha + ^{100}$Sn systems, where the nuclear term is the double-folding potential with \mbox{$\mathrm{M3Y} + c_{\mathrm{sat}} \delta(s)$} interaction (solid red line). For each \mbox{$\alpha$ + core} system, the respective potential composed by the nuclear term of \mbox{(1 + Gaussian)$\times$(W.S.~+ W.S.$^3$)} shape is shown (dashed blue line).}
\label{Fig_potentials}
\end{figure*}

The present authors introduced the use of the \linebreak[4] \mbox{(1 + Gaussian)$\times$(W.S.~+ W.S.$^3$)} nuclear potential to investigate the $\alpha$ + core structure in $^{46,54}$Cr \cite{SM2017}. Subsequently, the \mbox{(1 + Gaussian)$\times$(W.S.~+ W.S.$^3$)} shape was successfully applied to describe the properties of the set \{$^{20}$Ne, $^{44}$Ti, $^{94}$Mo, $^{104}$Te, $^{212}$Po\} \cite{SMB2019} and several even-even nuclei of the $22 \leq Z \leq 42$ region \cite{SM2021}. The recent work of J.~Jia, Y.~Qian, and Z.~Ren \cite{JQR2021} shows that the \mbox{(1 + Gaussian)$\times$(W.S.~+ W.S.$^3$)} potential shape can be successfully applied to heavy nuclei other than $^{212}$Po, including the analysis of different cluster structures above $^{208}$Pb. Such a potential has the \mbox{(1 + Gaussian)} factor which favors the correct reproduction of the $0^{+}$ ground state (g.s.)~and a better description of the $0^{+} \rightarrow 2^{+}$ spacing compared to the original \mbox{W.S.~+ W.S.$^3$} potential. The \mbox{(1 + Gaussian)} factor produces a depth increase on the \mbox{W.S.~+ W.S.$^3$} potential in a region restricted to \mbox{$r \lessapprox 1$ fm} (see Fig.~\ref{Fig_potentials}), acting more strongly on the \mbox{$0^{+} \rightarrow 2^{+}$} spacing; with the inclusion of the centrifugal potential ($L > 0$), the \mbox{(1 + Gaussian)} factor acts more weakly on the levels above $2^{+}$. So, the question arises whether the effect produced by the \mbox{(1 + Gaussian)} factor could be described in terms of the effective nucleon-nucleon ($NN$) interaction, since the double-folding potentials used in previous works do not present such a feature. Furthermore, the reproduction of the \mbox{(1 + Gaussian)} effect on the double-folding potential can provide a more in-depth interpretation of the \mbox{(1 + Gaussian)$\times$(W.S.~+ W.S.$^3$)} potential beyond its properties observed phenomenologically.

The \mbox{(1 + Gaussian)} factor range ($r \lessapprox 1$ fm) corresponds to the $\alpha$-core separation range in which the central regions of $\alpha$ and core overlap. In the central region of the nucleus (cluster or core), it is known that the nuclear matter density approaches or even exceeds the so-called \emph{saturation density}, $\rho_{\mathrm{sat}} \approx 0.17$ fm$^{-3}$. So, it is coherent to associate the \mbox{(1 + Gaussian)} factor range region with the overlap of $\alpha$ and core saturation regions. It should be considered that the superposition of cluster and core mass distributions is an intrinsic property of the double-folding model, as discussed in previous studies on \mbox{$\alpha$ + nucleus} collisions and heavy ion collisions \cite{KSO1997,Y2016}. The assumption of overlap of cluster and core saturation regions can lead to a discussion on the Pauli blocking effect at small internuclear separations \cite{SOV2001}; however, we restrict this work to a phenomenological approach focused on the systematic description of experimental data.

With such a premise, this work proposes to find a modification in the effective $NN$ interaction which produces an effect similar to the \mbox{(1 + Gaussian)} factor on the nuclear double-folding potential. In principle, this proposal could be applied to any effective $NN$ interaction form used in previous studies; starting from a simple option, the well-known M3Y-Reid interaction is chosen for this investigation. The M3Y-Reid effective $NN$ interaction is based on the $G$-matrix elements of the Reid $NN$ potential, and was originally developed as the real part of the optical potential for the description of heavy-ion scattering \cite{SL1979}. Differently from such use, the present work aims to develop a modified $NN$ interaction for the satisfactory reproduction of energy bands without the use of $L$-dependence, $B(E2)$ transition rates, $\alpha$-widths and $\alpha$-decay half-lives. Therefore, it is natural that the $\alpha$ + core potential proposed in this work presents limitations if applied in the description of $\alpha$ scattering data.

The article is developed as follows: in Section \ref{Sec:model}, a detailed discussion is made on the $\alpha$ + core potential adopted for the set \{$^{20}$Ne, $^{44}$Ti, $^{94}$Mo, $^{104}$Te, $^{212}$Po\}, the proposed $NN$ effective interaction, the matter density functions for $\alpha$ and core, the quantum number $G_{\mathrm{g.s.}}$ applied to the g.s.~band of each nucleus, and the similarity between the \mbox{(1 + Gaussian)$\times$(W.S.~+ W.S.$^3$)} potential and the double-folding potential with the proposed modification; in Section \ref{Sec:results}, the results obtained for energy levels, $B(E2)$ transition rates, $\alpha$-widths, half-lives, rms intercluster separations, and the reduced $\alpha$-widths are shown, including a comparison of the $^{104}$Te reduced $\alpha$-widths and the same quantities obtained for $^{94}$Mo and $^{212}$Po; conclusions are presented in Section \ref{Sec:conclusions}.

\section{$\alpha $-cluster model}
\label{Sec:model}

\subsection{$\alpha$ + core potential}
\label{Subsec:potential}

The properties of the nucleus are described in terms of an \mbox{$\alpha$ + core} system from the viewpoint of LPM, without considering internal excitations of $\alpha$-cluster and core. The \mbox{$\alpha$ + core} interaction is described through the local potential

\begin{equation}
V(r) = V_N(r) + V_C(r)
\end{equation}

\noindent containing the nuclear and Coulomb terms. The Coulomb potential $V_C(r)$ is that of an $\alpha$-particle interacting with an uniformly charged spherical core of radius $R$,

\begin{equation}
V_{C}(r)=\left\{ \begin{array}{ll}
{\displaystyle \left(\frac{e^{2}}{4\pi\varepsilon_{0}}\right)\frac{Z_{\alpha}\, Z_{\mathrm{c}}}{2R}\left(3-\frac{r^{2}}{R^{2}}\right)} & \mathrm{for}\quad r<R\\[14pt]
{\displaystyle \left(\frac{e^{2}}{4\pi\varepsilon_{0}}\right)\frac{Z_{\alpha}\, Z_{\mathrm{c}}}{r}} & \mathrm{for}\quad r\geq R\end{array}\right.
\label{Vcoulb}
\end{equation}

\noindent where $e^{2}/(4\pi\varepsilon_{0}) = 1.44 \; \mathrm{MeV} \cdot \mathrm{fm}$, $Z_{\alpha}$ and $Z_{\mathrm{c}}$ are the charge numbers of $\alpha$-cluster and core, respectively. For the parameter $R$, the same values used successfully in the $\alpha$ + core potential of Ref.~\cite{SMB2019} are employed, since that potential also includes a Coulomb term based on a uniformly charged sphere and is applied for the same set of nuclei. The nuclear potential $V_N(r)$ is given by the double-folding form

\begin{equation}
V_{N}(r)=\lambda\iint\rho_{\alpha}(\vec{r}_{\alpha})\rho_{c}(\vec{r}_{c})v_{NN}(\vec{s})d\vec{r}_{\alpha}d\vec{r}_{c}
\end{equation}

\noindent where $\vec{s} = \vec{r}-\vec{r}_{c}+\vec{r}_{\alpha}$, $\lambda$ is the intensity parameter, $v_{NN}(\vec{s})$ is the effective $NN$ interaction, $\rho_{\alpha}(\vec{r}_{\alpha})$ and $\rho_{c}(\vec{r}_{c})$ are the nucleon densities of the $\alpha$-cluster and core, respectively.

As discussed in Section \ref{Sec:introduction}, it is required that the nuclear potential $V_N(r)$ be such that it becomes more strongly attractive when the saturation regions of $\alpha$-cluster and core overlap, which occurs in a region restricted to $r \lessapprox 1$ fm. Therefore, in the systematics adopted here, we assume that both $\alpha$ and core have a central mass region where $\rho \geq 0.17$ fm$^{-3}$. The proposed shape for $V_N(r)$ can be produced by a suitable modification in the effective $NN$ interaction $v_{NN}(s)$. In this work, a variation of the traditional M3Y-Reid interaction is employed:

\begin{multline}
v_{NN}(s)  = {} \\[1pt]
V_{1}\frac{\exp(-\mu_{1}s)}{\mu_{1}s}+V_{2}\frac{\exp(-\mu_{2}s)}{\mu_{2}s}+\hat{J}_{00}\delta(s)+c_{\mathrm{sat}}\delta(s) \;.
\label{Eq_NN_interaction}
\end{multline}

\noindent The first three terms of eq.~\eqref{Eq_NN_interaction} are those present in the original form of M3Y interaction, while the term $c_{\mathrm{sat}}\delta(s)$ acts as an attractive interaction \emph{only between the saturation regions of $\alpha$-cluster and core}. This condition is expressed by:

\[
c_{\mathrm{sat}} = 0 \; \; \mathrm{for} \; \rho_{\alpha}(r_{\alpha}) < 0.17 \: \mathrm{fm}^{-3} \; \mathrm{or} \; \rho_{c}(r_{c}) < 0.17 \: \mathrm{fm}^{-3} \;;
\]

\[
c_{\mathrm{sat}} < 0 \; \; \mathrm{for} \; \rho_{\alpha}(r_{\alpha}) \geq 0.17 \: \mathrm{fm}^{-3} \; \mathrm{and} \; \rho_{c}(r_{c}) \geq 0.17 \: \mathrm{fm}^{-3} \;.
\]

\noindent The values for the parameters $V_1$, $V_2$, $\mu_1$, and $\mu_2$ are those applied to the M3Y-Reid interaction: $V_1 = 7999$ MeV, $V_2 = 2134$ MeV, $\mu_1 = 4$ fm$^{-1}$, and $\mu_2 = 2.5$ fm$^{-1}$. In the original M3Y form, $\hat{J}_{00}$ is a factor smoothly dependent on the bombarding energy $E_{\alpha}$ of the $\alpha$-particle in the scattering process \cite{SL1979}; however, several energy levels calculated for $^{20}$Ne, $^{44}$Ti, and $^{94}$Mo are below the \mbox{$\alpha$ + core} threshold, making such a dependence on $E_{\alpha}$ inappropriate in this context. Therefore, we adopted a fixed value $\hat{J}_{00} = - 276$ MeV, which corresponds to $E_{\alpha} = 0$ in the original M3Y-Reid form. The parameters $\lambda$ and $c_{\mathrm{sat}}$ are fitted to precisely reproduce the $0^{+}$ and $4^{+}$ levels of the $^{20}$Ne, $^{44}$Ti, $^{94}$Mo, and $^{212}$Po g.s.~bands\footnote{Details on the values of $\lambda$, $c_{\mathrm{sat}}$, and $R$ adopted for $^{104}$Te are discussed in Section \ref{Sec:results}.}. The values of $R$, $\lambda$ and $c_{\mathrm{sat}}$ applied to the g.s.~bands are shown in Table \ref{Table_parameters}; in addition, this table shows values of the ratio $r_{0 \, \mathrm{Coul.}} = R / A_{T}^{1/3}$, where $A_{T}$ is the mass number of the total nucleus. Note the small variation of $r_{0 \, \mathrm{Coul.}}$ for the set of nuclei, in the range of approx.~1.2$-$1.3 fm, which is compatible with the relation $R = 1.224 \, A_{T}^{1/3}$ (fm) obtained in Ref.~\cite{SMB2019}. For the negative parity bands, $\lambda$ is slightly modified for the correct reproduction of the experimental $1^{-}$ levels of $^{20}$Ne ($E_x = 5.7877$ MeV) and $^{44}$Ti ($E_x = 6.220$ MeV), and $(5)^{-}$ level of $^{94}$Mo ($E_x = 2.61057$ MeV), while the values of $R$ and $c_{\mathrm{sat}}$ are those in Table \ref{Table_parameters} (details in Section \ref{Sec:results}).

\begin{table}
\caption{Values of the parameters $\lambda$, $c_{\mathrm{sat}}$, and $R$, the ratio $r_{0 \, \mathrm{Coul.}} = R / A_{T}^{1/3}$, and the quantum number $G_{\mathrm{g.s.}}$ applied to the g.s.~bands of $^{20}$Ne, $^{44}$Ti, $^{94}$Mo, $^{104}$Te, and $^{212}$Po.}
\label{Table_parameters}
\begin{center}
\begin{tabular}{cccccc}
\hline
&  &  &  &  &  \\[-7pt] 
      &      &     &  $c_{\mathrm{sat}}$ & $R$ & $r_{0 \, \mathrm{Coul.}}$ \\[1pt]
Nucleus & $G_{\mathrm{g.s.}}$  & $\lambda$  &  (MeV) & (fm) & (fm) \\[2pt] \hline
&  &  &  &  &  \\[-6pt]
$^{20}$Ne  &  8   &  0.88720  &  $-2020$  & 3.272 & 1.205 \\
$^{44}$Ti  &  12   &  0.79450  &  $-3200$  &  4.551 & 1.289 \\
$^{94}$Mo  &  16   &  0.71560 &  $-5650$  &  5.783 & 1.272  \\
$^{104}$Te  &  16   &  0.65310  &  $-4870$  &  5.713 & 1.215 \\
$^{212}$Po  &  24   &  0.79005  &  $-5370$  &  7.018 & 1.177 \\[2pt] \hline
\end{tabular}
\end{center}
\end{table}

In Table \ref{Table_parameters}, it is seen the parameter $\lambda$ varies in the range of 0.6$-$0.9, being below the range of 1.1$-$1.4 applied in similar calculations with the double-folding nuclear potential with DDM3Y interaction \cite{O1995,HMS1994,M2017,AMA1996}. The use of M3Y interaction generates an unnormalized nuclear potential deeper than that obtained with the DDM3Y interaction, as discussed by Kobos {\it et al.}~\cite{KBH1982}, and this behavior extends to the \mbox{M3Y + $c_{\mathrm{sat}}\delta(s)$} interaction; therefore, there is the need for a smaller intensity parameter ($\lambda < 1.0$) to reproduce the energy bands of the nuclei considered.

The calculation of the $\alpha$ + core potential depends on determining the density functions $\rho_{\alpha}(r_{\alpha})$ and $\rho_{c}(r_{c})$; moreover, it is necessary to define the saturation region $r_{\alpha(c)} \leq R_{\mathrm{sat}}$, where $\rho(r_{\alpha(c)}) \geq 0.17$ fm$^{-3}$, for the density functions of $\alpha$ and core. The radius $R_{\mathrm{sat}}$ is determined from the parameter $\sigma$ of the (1 + Gaussian)$\times$(W.S.~+ W.S.$^3$) nuclear potential, as detailed in Subsection \ref{Subsec:saturation}. For the $\alpha$-cluster density, the 3-parameter Fermi function

\begin{equation}
\rho_{\alpha}(r_{\alpha})=\rho_{\alpha 0}\left(1+\frac{wr_{\alpha}^{2}}{c^{2}}\right)\left/ \left[1+\exp\left(\frac{r_{\alpha}-c}{z}\right)\right]\right.
\label{eq:ro_alfa}
\end{equation}

\noindent is used, as proposed in Ref.~\cite{FMR1967} for the $\alpha$-particle charge distribution. The original parameter values $c = 1.01$ fm and $w = 0.445$ are applied, while $\rho_{\alpha 0}$ and $z$ are fitted to normalize the total mass to $A_{\alpha} = 4$ and reproduce the saturation radius $R_{\mathrm{sat}}$ in the $\alpha$-cluster mass distribution. For the core density, the Woods-Saxon type function is applied:

\begin{equation}
\rho_{c}(r_{c})=\rho_{c0}\left/ \left[1+\exp\left(\frac{r_{c} - R_{c}}{a}\right)\right]\right.
\label{eq:WSaxon}
\end{equation}

\noindent For the diffuseness parameter, the typical value $a = 0.55$ fm is used, while $\rho_{c0}$ and $R_c$ are fitted to normalize the core mass to $A_c$ nucleons and reproduce the saturation radius $R_{\mathrm{sat}}$ in the core mass distribution. The values of $\rho_{\alpha 0}$, $z$, $\rho_{c0}$, and $R_c$ applied to the $\alpha$ + core systems are shown in Table \ref{Table_param_density}; in addition, this table shows values of the ratio $r_{0} = R_{c} / A_{c}^{1/3}$. There is a small variation of $r_{0}$ for the set of nuclei, in the range of approx.~1.0$-$1.1 fm, which is compatible with the relation $R_{c} = r_{0} \, A_{c}^{1/3}$ (fm) expected for the core matter distribution.

\begin{table}
\caption{Values applied to the variable parameters of $\alpha$-cluster ($\rho_{\alpha 0}$ and $z$) and core ($\rho_{c 0}$ and $R_c$) density functions, and the ratio $r_{0} = R_{c} / A_{\mathrm{c}}^{1/3}$.}
\label{Table_param_density}
\begin{center}
\begin{tabular}{cccccc}
\hline
&  &  &  &  &  \\[-7pt] 
      &   $\rho_{\alpha 0}$   &  $z$   &  $\rho_{c 0}$ & $R_c$ & $r_{0}$ \\[1pt]
System & (fm$^{-3}$)  & (fm)  &  (fm$^{-3}$) & (fm) & (fm) \\[2pt] \hline
&  &  &  &  &  \\[-6pt]
$\alpha + ^{16}$O  &  0.19170   &  0.3677  &  0.17503  & 2.4400 & 0.968 \\
$\alpha + ^{40}$Ca  &  0.19046   &  0.3690  &  0.17059  &  3.5657 & 1.043 \\
$\alpha + ^{90}$Zr  &  0.19014   &  0.3693 &  0.17006  &  4.8198 & 1.076 \\
$\alpha + ^{100}$Sn  &  0.19082   &  0.3686 &  0.17004  &  5.0062 & 1.079 \\
$\alpha + ^{208}$Pb  &  0.19094   &  0.3685 &  0.17000  &  6.4850 & 1.095 \\[2pt] \hline
\end{tabular}
\end{center}
\end{table}

\subsection{Saturation radius $R_{\mathrm{sat}}$}
\label{Subsec:saturation}

The saturation radius $R_{\mathrm{sat}}$ is estimated from the $\sigma$ parameter of the \mbox{(1 + Gaussian)$\times$(W.S.~+ W.S.$^3$)} nuclear potential. The factor (1 + Gaussian) has the form $1 + \lambda \exp(-r^2/\sigma^2)$, where $\lambda = 0.14$ is one of the fixed parameters of the \mbox{(1 + Gaussian)$\times$(W.S.~+ W.S.$^3$)} potential; this factor approaches a constant value = 1 from $r \approx 3\sigma/\sqrt{2}$. From the viewpoint of the double-folding model, the radius $r \approx 3\sigma/\sqrt{2}$ can be interpreted as the cluster-core separation in which the $\alpha$-cluster and core saturation regions begin to overlap, so that the $c_{\mathrm{sat}} \delta(s)$ term of the $NN$ interaction becomes relevant in $r \lessapprox 3\sigma/\sqrt{2}$. Such an interpretation is expressed by

\begin{equation}
R_{\mathrm{sat}\, \alpha} + R_{\mathrm{sat}\, c} = \frac{3\sigma}{\sqrt{2}} \;,
\end{equation}

\noindent where $R_{\mathrm{sat}\, \alpha}$ and $R_{\mathrm{sat}\, c}$ are the saturation radii of $\alpha$-cluster and core, respectively. Assuming that $R_{\mathrm{sat}\, \alpha}$ and $R_{\mathrm{sat}\, c}$ are similar in range, we consider both as a single parameter $R_{\mathrm{sat}}$, resulting in

\begin{equation}
R_{\mathrm{sat}} = \frac{3\sigma}{2\sqrt{2}} \;.
\end{equation}

\begin{figure*}
\begin{center}
\includegraphics[scale=0.78]{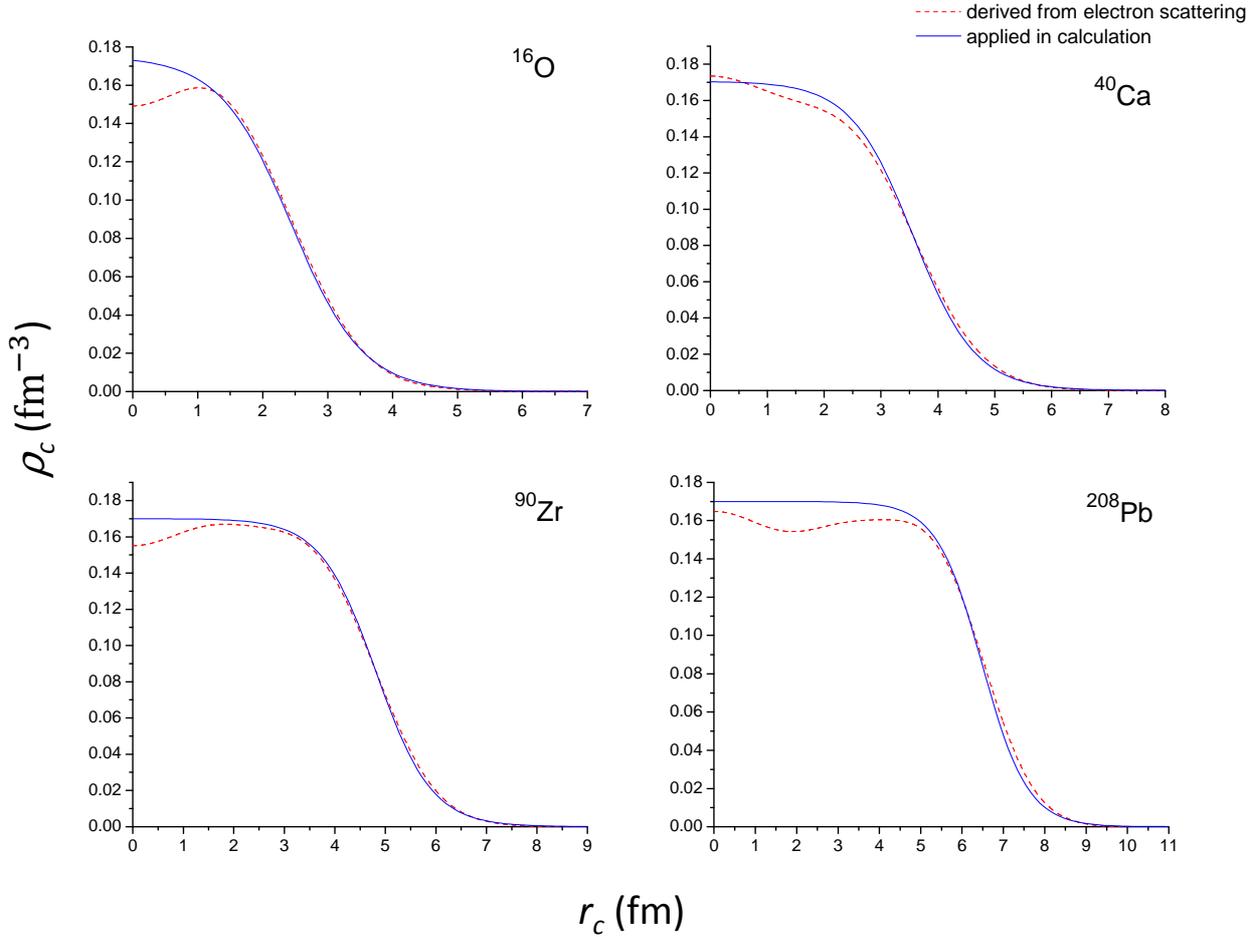}
\end{center}
\caption{Nuclear densities $\rho_{c}(r_{c})$ applied to $^{16}$O, $^{40}$Ca, $^{90}$Zr, and $^{208}$Pb in the double-folding calculations (solid blue lines). For comparison, the respective charge distributions normalized by the $A/Z$ factor are shown (dashed red lines). The charge distributions are deduced from elastic electron scattering and Fourier-Bessel analysis \cite{Densities}.}
\label{Fig_densities}
\end{figure*}

Using the estimated $R_{\mathrm{sat}}$ values for the $\alpha + ^{16}$O, $\alpha + ^{40}$Ca, $\alpha + ^{90}$Zr, $\alpha + ^{100}$Sn, and $\alpha + ^{208}$Pb systems, the respective density functions $\rho_{\alpha}(r_{\alpha})$ and $\rho_{c}(r_{c})$ are determined. Figure \ref{Fig_densities} shows the comparison of the functions $\rho_{c}(r_{c})$ obtained for $^{16}$O, $^{40}$Ca, $^{90}$Zr, and $^{208}$Pb with the respective charge distributions deduced from elastic electron scattering \cite{Densities}, normalized by the $A/Z$ factor. The proximity of the functions $\rho_{c}(r_{c})$ and the respective normalized charge distributions is clearly noted, mainly in the nuclear surface region.

\subsection{Quantum number $G_{\mathrm{g.s.}}$}
\label{Subsec:band_number}

The $\alpha$-cluster nucleons must lie in shell-model orbitals outside the core. This restriction is defined through the global quantum number $G = 2N + L$, where $N$ is the number of internal nodes in the radial wave function and $L$ is the orbital angular momentum; thus, the restriction is expressed by $G \geq G_{\mathrm{g.s.}}$, where $G_{\mathrm{g.s.}}$ corresponds to the g.s.~band. The numbers $G_{\mathrm{g.s.}} = 8$ and 12 are applied to the g.s.~bands of $^{20}$Ne and $^{44}$Ti, respectively, and $G_{\mathrm{g.s.}} = 16$ is applied to $^{94}$Mo and $^{104}$Te, according to the Wildermuth condition \cite{WT1977}. $G_{\mathrm{g.s.}} = 24$ is applied to the g.s.~band of $^{212}$Po, as this number provides a satisfactory reproduction of the experimental energy levels from 0$^{+}$ to 8$^{+}$ using the double-folding potential, including the compression trend of the energy spacing from the $4^{+}$ level. The number $G_{\mathrm{g.s.}} = 24$ has been suitably used in previous work as \cite{BMP1991,BMP1992} to describe the $\alpha$-decay half-life of $^{212}$Po, using other $\alpha$ + core potential shapes.

\subsection{Comparison of double-folding and \mbox{(1 + Gaussian)$\times$(W.S.~+ W.S.$^3$)} potentials}

Using the methodology proposed in subsections \ref{Subsec:potential}, \ref{Subsec:saturation}, and \ref{Subsec:band_number}, the $\alpha$ + core potential was determined for analysis of $^{20}$Ne, $^{44}$Ti, $^{94}$Mo, $^{104}$Te, and $^{212}$Po. A comparison of the $\alpha$ + core potentials obtained with the double-folding and \mbox{(1 + Gaussian)$\times$(W.S.~+ W.S.$^3$)} nuclear terms is shown in Figure \ref{Fig_potentials} for $^{20}$Ne, $^{44}$Ti, $^{94}$Mo, and $^{104}$Te. Note that the two potential forms are similar for the analyzed nuclei, including the $r \leq 1$ fm region where both become more intensely attractive. Therefore, it is shown that the effect of the (1 + Gaussian) factor can be reproduced in the double-folding potential with the proposed modification in the effective $NN$ interaction. For $^{212}$Po, such a comparison is not presented since there is a difference between the $G_{\mathrm{g.s.}}$ number applied in this work with the double-folding potential ($G_{\mathrm{g.s.}} = 24$) and that applied with the \mbox{(1 + Gaussian)$\times$(W.S.~+ W.S.$^3$)} potential in Ref.~\cite{SMB2019} \linebreak[4] ($G_{\mathrm{g.s.}} = 20$).

\section{Results}
\label{Sec:results}

\begin{figure*}
\begin{center}
\includegraphics[scale=0.78]{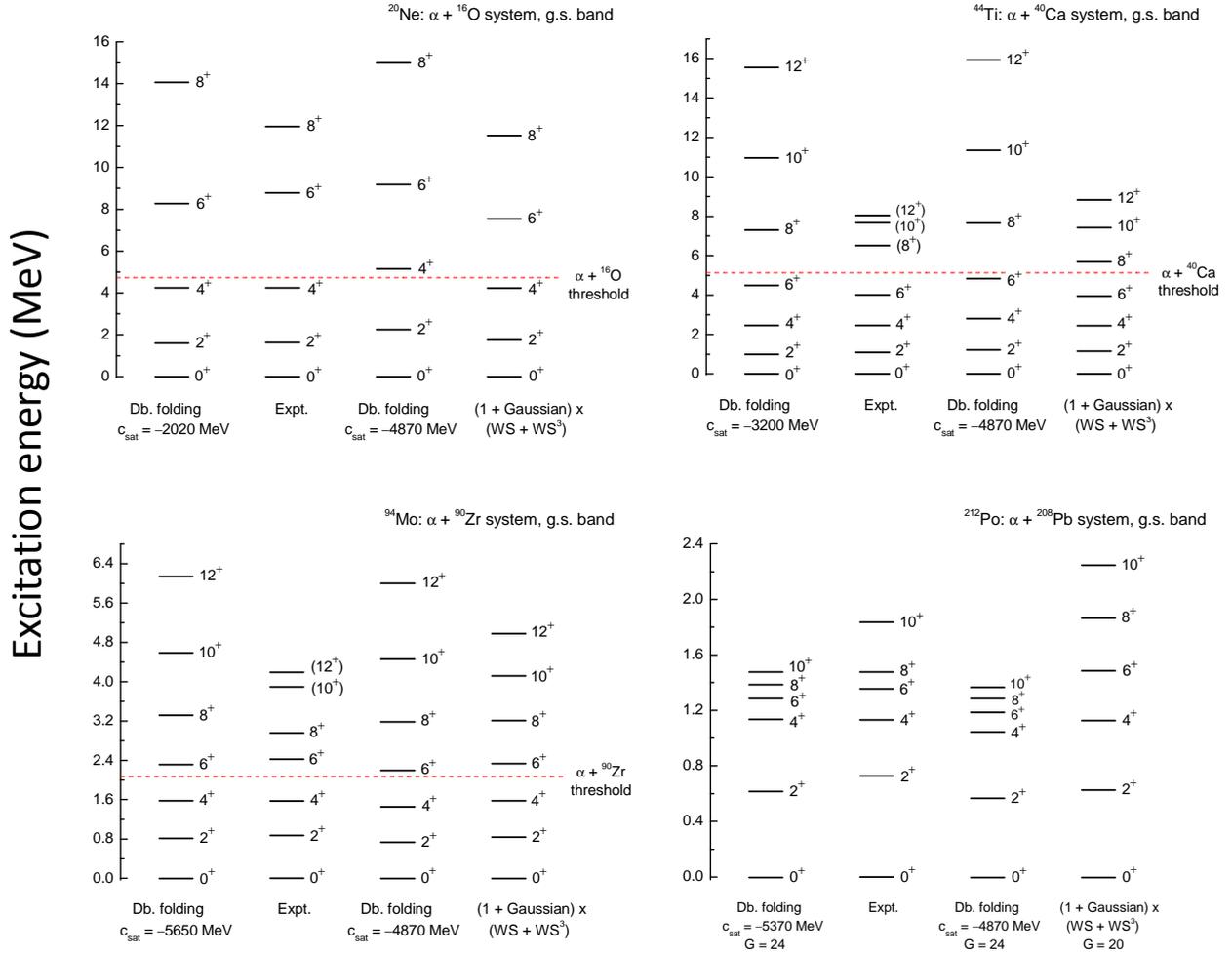}
\end{center}
\caption{Ground state bands calculated for the $\alpha + ^{16}$O ($G_{\mathrm{g.s.}} = 8$), $\alpha + ^{40}$Ca ($G_{\mathrm{g.s.}} = 12$), $\alpha + ^{90}$Zr ($G_{\mathrm{g.s.}} = 16$), and $\alpha + ^{208}$Pb ($G_{\mathrm{g.s.}} = 24$) systems using the double-folding nuclear potential with $\mathrm{M3Y} + c_{\mathrm{sat}} \delta(s)$ interaction. For each nucleus, the g.s.~band is calculated firstly with $c_{\mathrm{sat}}$ fitted for the correct reproduction of the $0^{+}$ and $4^{+}$ states, and secondly, with $c_{\mathrm{sat}} = -4870$ MeV. The fixed value of the intensity parameter ($\lambda$) applied for each $\alpha$ + core system is shown in Table \ref{Table_parameters}. In addition, the g.s.~bands calculated with the \mbox{(1 + Gaussian)$\times$(W.S.~+ W.S.$^3$)} nuclear potential \cite{SMB2019} are shown for comparison.}
\label{Fig_gs_bands}
\end{figure*}

The g.s.~bands of $^{20}$Ne, $^{44}$Ti, $^{94}$Mo, and $^{212}$Po are calculated with the $\alpha$ + core potential defined in \mbox{Section \ref{Sec:model}} and the parameter values in Table \ref{Table_parameters}. The resolution of Schr\"{o}dinger radial equation for the $\alpha$ + core relative motion allows to determine the energy levels of the system and respective radial wave functions. Figure \ref{Fig_gs_bands} shows the comparison of calculated and experimental energy levels. It is shown the $\alpha$ + core potential provides a good description of experimental levels from $0^{+}$ to $8^{+}$. For $L > 8$, the description of the experimental levels is rougher, as the double-folding potential produces an energy band with rotational behavior at higher spins for $^{20}$Ne, $^{44}$Ti, and $^{94}$Mo. However, it should be noted that the satisfactory results from $0^{+}$ to $8^{+}$ were obtained without using an $L$-dependent intensity parameter. As shown in Figure \ref{Fig_gs_bands}, the double-folding and \mbox{(1 + Gaussian)$\times$(W.S.~+ W.S.$^3$)} potentials produce similar bands for $^{20}$Ne, $^{44}$Ti, and $^{94}$Mo from $0^{+}$ to $8^{+}$, and a more visible difference at $L > 8$, where the \mbox{(1 + Gaussian)$\times$(W.S.~+ W.S.$^3$)} potential gives a better description of the experimental energy levels for $^{44}$Ti and $^{94}$Mo. In the case of $^{212}$Po, the double-folding potential produces an energy band compressed at $L > 4$, which favors a better reproduction of the experimental band compared to the \mbox{(1 + Gaussian)$\times$(W.S.~+ W.S.$^3$)} potential.

In a next calculation, a single value of the parameter $c_{\mathrm{sat}}$ is determined for the $^{20}$Ne, $^{44}$Ti, $^{94}$Mo, and $^{212}$Po nuclei. For this, $c_{\mathrm{sat}}$ is varied to minimize the quantity

\begin{equation}
S^2 = \sum (E_{\mathrm{expt}} - E_{\mathrm{calc}})^2
\end{equation}

\noindent for the energy levels from $0^{+}$ to $8^{+}$ of the $^{20}$Ne, $^{44}$Ti, $^{94}$Mo, and $^{212}$Po g.s.~bands. Using this procedure, the value $c_{\mathrm{sat}} = -4870$ MeV is obtained. In Figure \ref{Fig_gs_bands}, it is seen the bands calculated with the fixed value $c_{\mathrm{sat}} = -4870$ MeV are still satisfactory in describing the levels from $0^{+}$ to $8^{+}$, showing it is possible to describe the four nuclei through an effective $NN$ interaction with fixed parameters.

\begin{figure*}
\begin{center}
\includegraphics[scale=0.78]{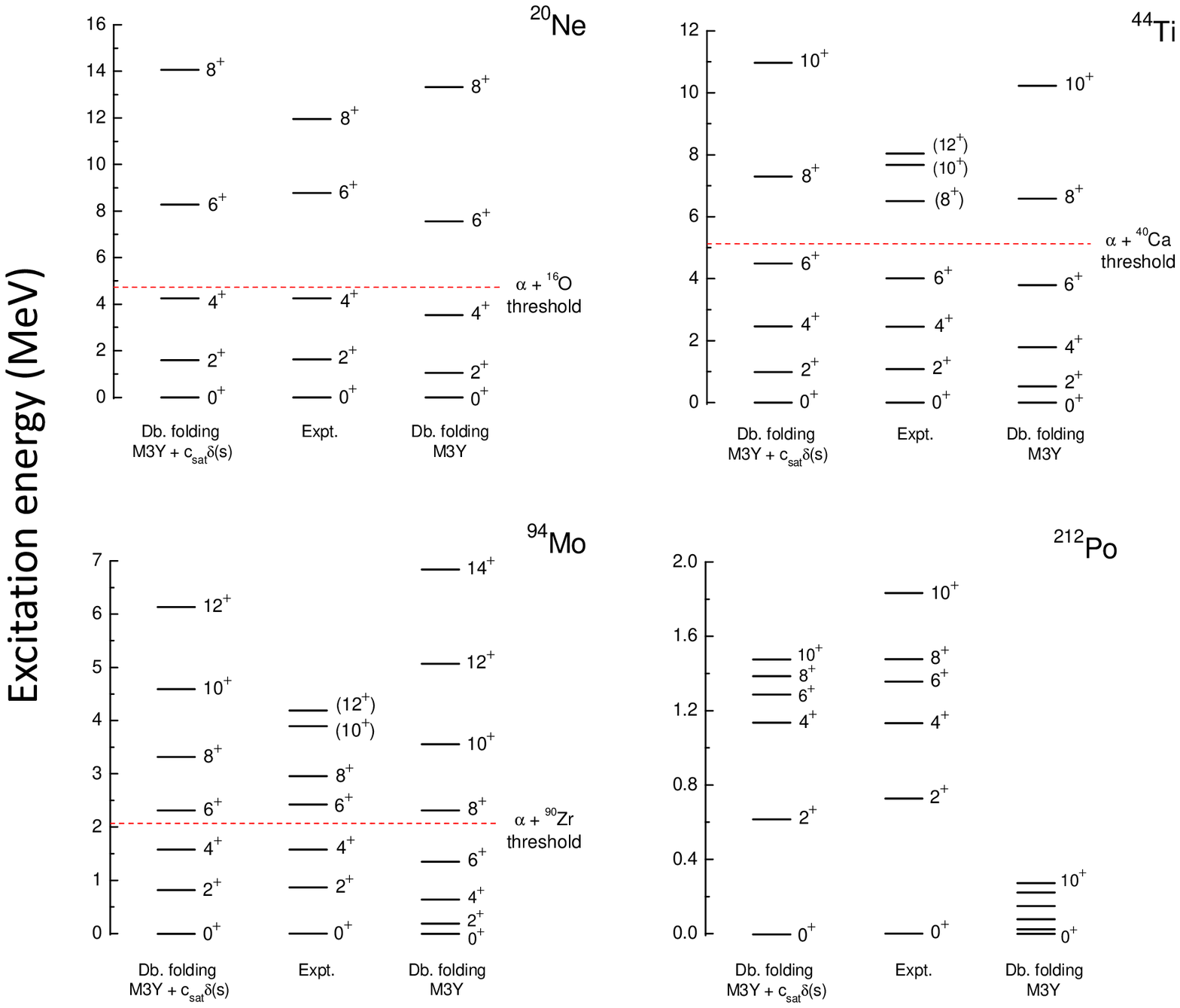}
\end{center}
\caption{Comparison of the ground state bands calculated for the $\alpha + ^{16}$O ($G_{\mathrm{g.s.}} = 8$), $\alpha + ^{40}$Ca ($G_{\mathrm{g.s.}} = 12$), $\alpha + ^{90}$Zr ($G_{\mathrm{g.s.}} = 16$), and $\alpha + ^{208}$Pb ($G_{\mathrm{g.s.}} = 24$) systems using the double-folding nuclear potential with the effective $NN$ interactions: M3Y and \mbox{$\mathrm{M3Y} + c_{\mathrm{sat}} \delta(s)$}. For each nucleus, $c_{\mathrm{sat}}$ is fitted for the correct reproduction of the $0^{+}$ and $4^{+}$ states.}
\label{Fig_M3Y}
\end{figure*}

It is interesting to compare the g.s.~bands calculated with and without the $c_{\mathrm{sat}} \delta(s)$ term of the effective $NN$ interaction, to verify its level of importance in reproducing the experimental levels. Figure \ref{Fig_M3Y} shows the g.s.~bands of $^{20}$Ne, $^{44}$Ti, $^{94}$Mo, and $^{212}$Po calculated with the effective $NN$ interactions: M3Y-Reid (applying $\hat{J}_{00} = - 276$ MeV) and the $v_{NN}(s)$ interaction of eq.~\eqref{Eq_NN_interaction}. In the case of $^{20}$Ne and $^{44}$Ti, it is seen that the contribution of the $c_{\mathrm{sat}} \delta(s)$ term is relevant for a better reproduction of the experimental levels from $0^{+}$ to $4^{+}$. However, in the case of $^{94}$Mo, the contribution of the $c_{\mathrm{sat}} \delta(s)$ term is more significant, providing a better reproduction of the experimental levels from $0^{+}$ to $8^{+}$ and canceling the rotational behavior produced with the M3Y interaction from $0^{+}$ to $8^{+}$. In the case of $^{212}$Po, the contribution of the $c_{\mathrm{sat}} \delta(s)$ term is evident. For this nucleus, the g.s.~band produced with the simple M3Y interaction is very compressed and incompatible with the experimental band; e.g., the calculated $10^{+}$ level is at $E_x = 0.273$ MeV, which is far below the respective experimental energy. With the inclusion of the $c_{\mathrm{sat}} \delta(s)$ term in the effective $NN$ interaction, the calculated band provides a good description of the experimental levels from $0^{+}$ to $8^{+}$, also reproducing the compression trend of the experimental band from the $4^{+}$ level. As the compression of the $^{212}$Po calculated band remains above $8^{+}$, only the experimental $10^{+}$ level is more roughly reproduced.

\begin{figure*}[t]
\begin{center}
\includegraphics[scale=0.86]{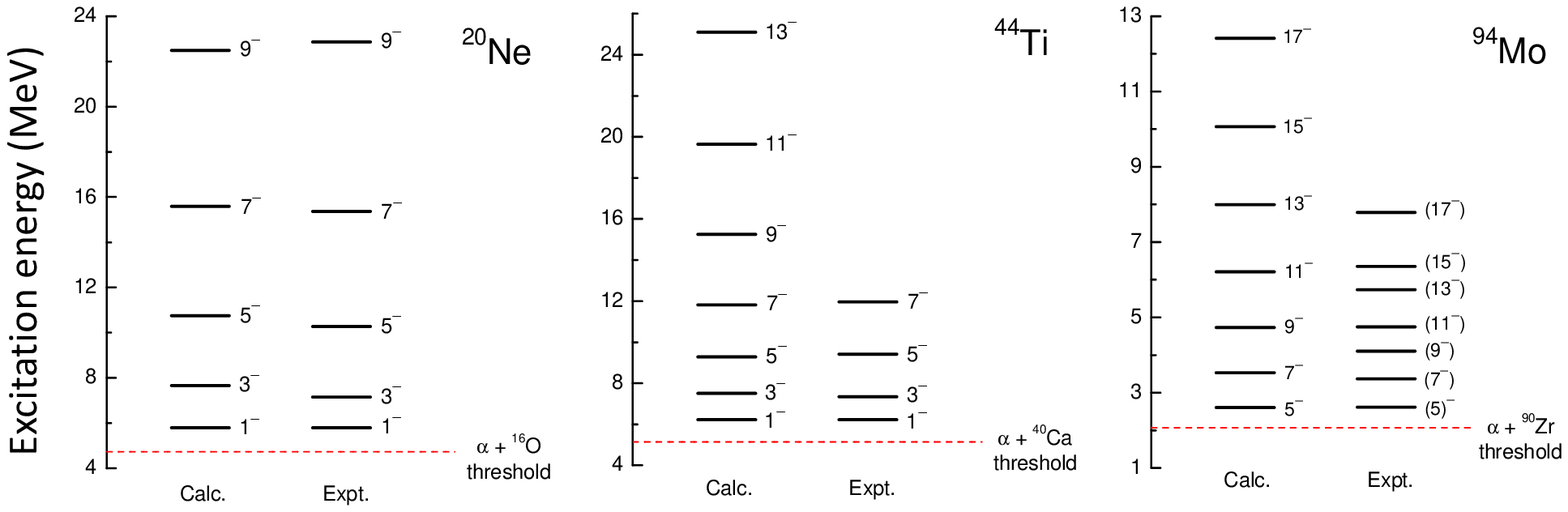}
\end{center}
\caption{Negative parity bands calculated for the $\alpha + ^{16}$O ($G = 9$), $\alpha + ^{40}$Ca ($G = 13$), and $\alpha + ^{ 90}$Zr ($G = 17$) systems through the double-folding nuclear potential with \mbox{$\mathrm{M3Y} + c_{\mathrm{sat}} \delta(s)$} interaction, in comparison with experimental energy levels \cite{ENSDF,ZHG2009}. The intensity parameter $\lambda$ is slightly modified for the correct reproduction of the $1^{-}$ levels of $^{20}$Ne and $^{44}$Ti, and (5)$^{-}$ of $^{94}$Mo (the $\lambda$ values are discussed in the text).}
\label{Fig_neg_bands}
\end{figure*}

The model also allows the calculation of excited bands of the $\alpha$ + core systems. Figure \ref{Fig_neg_bands} shows the calculated negative parity bands for $^{20}$Ne ($G = 9$), $^{44}$Ti ($G = 13$), and $^{94}$Mo ($G = 17$)\footnote{The $^{94}$Mo experimental negative parity band presents levels from (5)$^{-}$ ($E_x = 2.611$ MeV) to ($17^{-}$) ($E_x = 7.782$ MeV) connected by $\gamma$-transitions \cite{ZHG2009}; however, the excited state (5)$^{-}$ undergoes a $\gamma$-transition to the $4^{+}$ state ($E_x = 1.574$ MeV) of the $G = 16$ g.s.~band. As the excited state (5)$^{-}$ does not undergo a transition of $J \rightarrow J - 2$ type, the $G = 17$ band is assumed to be incomplete and the calculation of the corresponding theoretical band is made from $J^{\pi} = 5^{-}$.} compared to experimental energies. For calculating these bands, the values of the intensity parameter of $V_N(r)$ are: $\lambda = 0.9025$, 0.8050, and 0.7703 for $^{20}$Ne, $^{44}$Ti, and $^{94}$Mo, respectively. The relative variation of $\lambda$ from the g.s.~band to the negative parity band is very small for $^{20}$Ne (1.72 \%) and $^{44}$Ti (1.32 \%), showing the $\alpha$ + core potential is weakly dependent on the band number $G$ for these two nuclei. In the case of $^{94}$Mo, the relative variation of $\lambda$ is somewhat greater (7.64 \%), and the experimental negative parity band is more compressed compared to the calculated band; however, the calculation for $^{94}$Mo should be interpreted as exploratory, since the spins and parities of the considered experimental levels are uncertain.

\begin{figure}
\begin{center}
\includegraphics[scale=0.35]{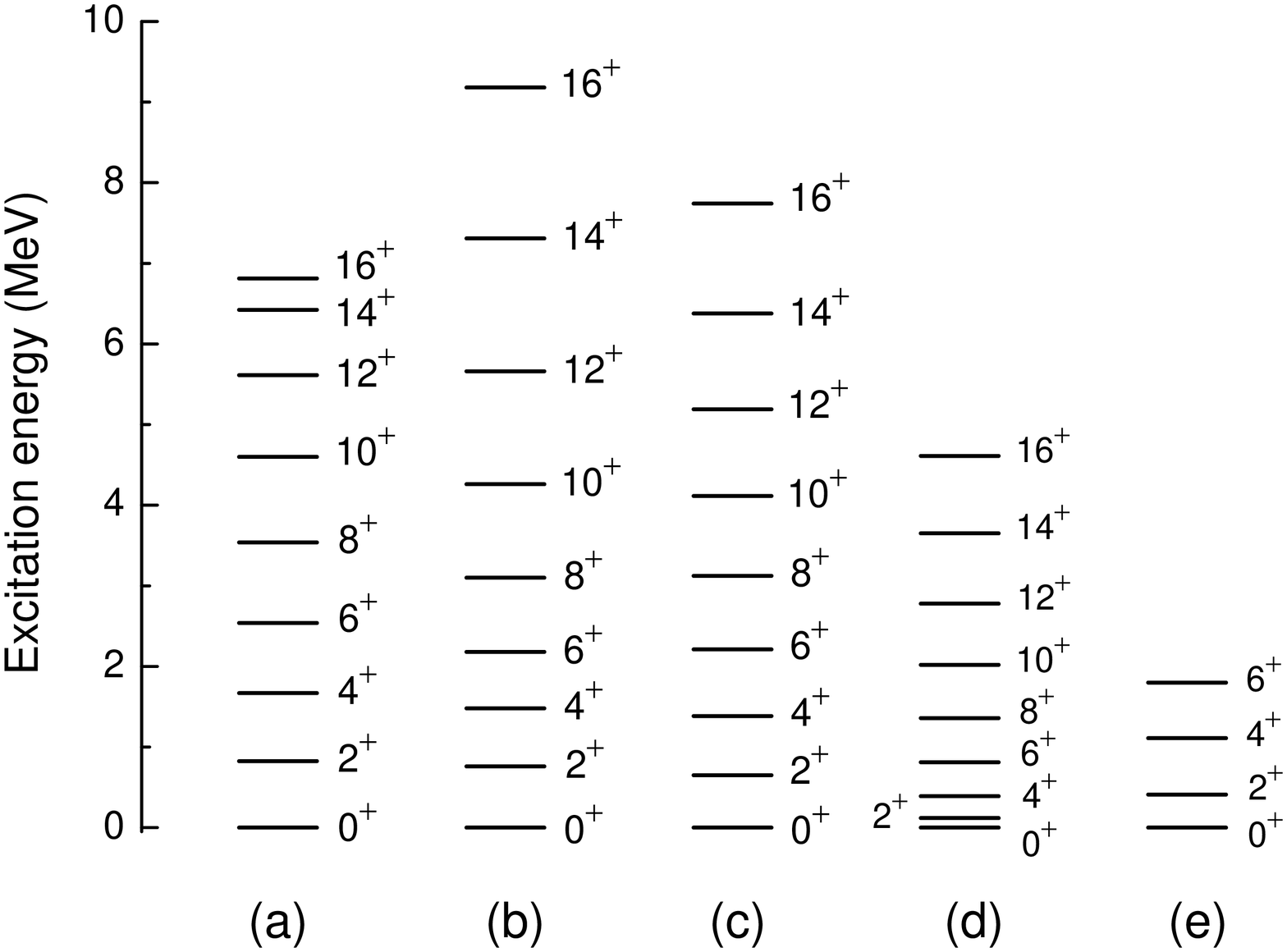}
\end{center}
\caption{Ground state band calculated for the $\alpha + ^{100}$Sn system ($G_{\mathrm{g.s.}} = 16$) in the cases: {\bf (a)} using the nuclear potential of \mbox{(1 + Gaussian)$\times$(W.S.~+ W.S.$^3$)} shape \cite{SMB2019}; {\bf (b)} using the double-folding nuclear potential with \mbox{$\mathrm{M3Y} + c_{\mathrm{sat}} \delta(s)$} effective interaction (this work); {\bf (c)} calculation of P.~Mohr \cite{M2007} using the double-folding nuclear potential with DDM3Y interaction and $L$-dependent intensity parameter; {\bf (d)} calculation of D.~Bai and Z.~Ren \cite{BR2018} using the double-folding nuclear potential with M3Y interaction and fixed intensity parameter; {\bf (e)} calculation of H.~Jiang {\it et al.} \cite{JQL2013} using the shell-model without isospin symmetry (SM1).}
\label{Fig_band_104Te}
\end{figure}

The g.s.~band calculated for $^{104}$Te is shown in Fig.~\ref{Fig_band_104Te}, case {\bf (b)}. In this calculation, the parameter $\lambda$ is fitted to reproduce the energy $E(0^{+}) = 5.22$ MeV used in Ref.~\cite{SMB2019} to describe the ground state of $^{104}$Te. In addition, $c_\mathrm{sat} = -4870$ MeV is adopted, since this value proved to be satisfactory for the set \{$^{20}$Ne, $^{44}$Ti, $^{94}$Mo, $^{212}$Po\}. As in the other nuclei, the value of $R$ used in Ref.~\cite{SMB2019} for the Coulomb potential is applied ($R = 5.713$ fm for $^{104}$Te). In Fig.~\ref{Fig_band_104Te}, the $^{104}$Te g.s.~band calculated in this work is compared with previous calculations in the excitation energy scale. Note that the energy levels calculated through the double-folding nuclear potential with \mbox{$\mathrm{M3Y} + c_{\mathrm{sat}} \delta(s)$} interaction are close to the respective $J^{+}$ levels obtained with the \mbox{(1 + Gaussian)$\times$(W.S.~+ W.S.$^3$)} nuclear potential from $0^{+}$ to $12^{+}$, indicating the similarity of the two potential shapes. Furthermore, there is a clear proximity (mainly from $0^{+}$ to $12^{+}$) between the band calculated with the \mbox{$\mathrm{M3Y} + c_{\mathrm{sat}} \delta(s)$} interaction and the band obtained by Mohr \cite{M2007} through the double-folding potential with DDM3Y interaction and $L$-dependent $\lambda$ parameter. The band calculated by D.~Bai and Z.~Ren \cite{BR2018} is more compressed and has strongly rotational behavior, which is produced by the double-folding potential with simple M3Y interaction. Also, it is seen the calculation of the present work generates energy spacings $2^{+} \rightarrow 4^{+}$ and $4^{+} \rightarrow 6^{+}$ similar to the shell-model calculation without isospin symmetry (SM1) by Jiang {\it et al.}~\cite{JQL2013}\footnote{Ref.~\cite{JQL2013} presents other shell-model calculations for the $^{104}$Te g.s.~band, such as the $SD$-pair approximation, the $SDGIK$-pair approximation and the shell-model calculation with isospin symmetry (SM2); however, such calculations produce results similar to the SM1 calculation shown in Fig.~\ref{Fig_band_104Te}.}. The evaluation of the calculated energy bands is limited by the absence of experimental data related to the excited states of $^{104}$Te; nevertheless, the comparison of the different $^{104}$Te theoretical bands indicates the $c_{\mathrm{sat}} \delta(s)$ term added to the effective $NN$ interaction produces an effect on energy levels similar to that generated by the $L$-dependent intensity parameter applied to the double-folding potential (except at the highest spin levels).

The $\alpha$-decay width ($\Gamma_{\alpha}$) of $^{104}$Te is calculated by the semi-classical approximation of Ref.~\cite{GK1987}, using an $\alpha$ preformation factor $P_{\alpha} = 1$; such an approximation is derived from the two-potential approach \cite{GK1987,QRN2011_JPG,QRN2011_PRC}, which is often applied in $\alpha$-decay studies. The $\alpha$-decay width obtained for the ground state at $E(0^{+}) = 5.22$ MeV is $\Gamma_{\alpha} = 1.0188 \times 10^{-13}$ MeV, and the corresponding $\alpha$-decay half-life, given by $T_{1/2,\alpha} = \hbar \ln 2 / \Gamma_{\alpha}$, is \mbox{$\approx 4.478$ ns}. This $T_{1/2,\alpha}$ value is compatible with the experimental measure of Auranen {\it et al.}~\cite{ASA2018} ($T_{1/2}^{\mathrm{exp}} < 18$ ns). By varying the parameter $\lambda$, it is verified that the energy range $5.044 \; \mathrm{MeV} < E(0^{+}) < 5.3$ MeV provides $T_{1/2,\alpha}$ values compatible with the experimental measure, and simultaneously, is compatible with the experimental measure $Q_{\alpha}^{\mathrm{exp}} = 5.1(2)$ MeV. In agreement with the calculation in Ref.~\cite{SMB2019}, the production of such results with $P_{\alpha} = 1$ corroborates the indication of superallowed $\alpha$-decay of $^{104}$Te by Auranen {\it et al.}

The $\alpha_0$-decay widths calculated for the $^{20}$Ne resonant states are shown in Table \ref{Table_Gamma_20Ne} in comparison with experimental values. The widths are calculated by the semi-classical approximation of Ref.~\cite{GK1987}, except for the $9^{-}$ level ($E_x = 22.87$ MeV), where the $\alpha_0$-width is calculated by \mbox{$\Gamma_{\alpha} = 2/(d\delta_L /dE)|_{E=E_r}$}, where $\delta_L$ is the scattering phase shift for a given $L$, and $E_R$ is the resonance energy. In this calculation, the parameter $\lambda$ is smoothly modified so that the experimental energy of the $J^\pi$ state is reproduced precisely. The $\Gamma_{\alpha_0}$ values calculated for the $6^{+}$ and $8^{+}$ levels of the g.s.~band have an order of magnitude above that observed for the respective experimental values; these results repeat a discrepancy found in some previous calculations on $^{20}$Ne \cite{BMP95,IMP2019,BR2021} with different $\alpha$ + core potentials. Concerning the $G = 9$ band, the calculated $\Gamma_{\alpha_0}$ values correctly reproduce the order of magnitude of the experimental $\alpha_0$-widths of $1^{-}$, $3^{-}$, $5^{-}$, $7^{-}$ ($E_x = 13.692$ MeV), and $9^{-}$ ($E_x = 17.430$ MeV) states. Such results corroborate the statement of previous works \cite{BMP95,BJM1995} in which the $7^{-}$ ($E_x = 13.692$ MeV) and $9^{-}$ ($E_x = 17.430$ MeV) states are better candidates for members of the $G = 9$ band of the $\alpha + ^{16}$O system; however, it is not excluded that both $7^{-}$ and both $9^{-}$ experimental levels in Table \ref{Table_Gamma_20Ne} can be associated with the theoretical $7^{-}$ and $9^{-}$ states of the $G = 9$ band.

\begin{table}
\begin{threeparttable}
\caption{Calculated $\alpha_0$-widths ($\Gamma_{\alpha_0}^{\mathrm{calc.}}$) for the resonant states of the ground state band ($G = 8$) and negative parity band ($G = 9$) of $^{20}$Ne in comparison with experimental data ($\Gamma_{\alpha_0}^{\mathrm{expt.}}$) \cite{TCK1998}. An $\alpha$ preformation factor $P_{\alpha} = 1$ is considered.}
\label{Table_Gamma_20Ne}
\begin{center}
\begin{tabular}{lcccc}
\hline
&  &  &  &  \\[-7pt] 
	&   &  $E_x$ & $\Gamma_{\alpha_0}^{\mathrm{calc.}}$ &  $\Gamma_{\alpha_0}^{\mathrm{expt.}}$ \\[1pt]
$J^\pi $ & $G$ &  (MeV)   & (keV) & (keV) \\ [3pt] \hline
 &  &  &  &  \\[-6pt]
$6^{+}$  &  8  &  8.7776  &  1.08  &  $0.11 \pm 0.02$  \\
$8^{+}$  &  8  &  11.951  &  0.74  &  $0.035 \pm 0.010$  \\
$1^{-}$  &  9  &  5.7877  &  0.061  &  $0.028 \pm 0.003$  \\
$3^{-}$  &  9  &  7.1563  &  18.5  &  $8.2 \pm 0.3$  \\
$5^{-}$  &  9  &  10.262  &  302  &  $145 \pm 40$  \\
$7^{-}$\tnote{*}  &  9  & 13.692  &  240 &  $158 \pm 15$  \\
$7^{-}$  &  9  &  15.366  &  890  &  $78 \pm 7$  \\
$9^{-}$\tnote{*}  &  9  & 17.430  &  37.2 &  $53 \pm 6$  \\
$9^{-}$  &  9  &  22.87   &  $1.36 \times 10^3$  &  $(0.225 \pm 0.040) \times 10^3$  \\[2pt] \hline
\end{tabular}
\begin{tablenotes}
\item [*] Energy levels not shown in Fig.~\ref{Fig_neg_bands}.
\end{tablenotes}
\end{center}
\end{threeparttable}
\end{table}

\begin{table*}
\begin{center}
\begin{threeparttable}
\caption{$\gamma$-decay widths ($\Gamma_{\gamma}$), $\alpha$-decay widths ($\Gamma_{\alpha}$), $\alpha$-branching ratios ($b_{\alpha}$), and half-lives ($T_{1/2}$) calculated for the states from $0^{+}$ to $8^{+}$ of the $^{212}$Po g.s.~band, in comparison with experimental data (\cite{ENSDF}, except where indicated). The values adopted for the internal conversion coefficient ($\alpha_{IC}$) are obtained from Ref.~\cite{KBT2008}. An $\alpha$ preformation factor $P_{\alpha} = 3.902 \, \%$ is applied.}
\label{Widths_212Po}
\begin{tabular}{ccccccccc}
\hline
&  &  &  &  &  &  &  &  \\[-7pt] 
       &  $E_x$ 	&   	 &   $\Gamma_{\gamma}(J\rightarrow J-2)$	&  $\Gamma_{\alpha}$   &  $b_{\alpha}^{\mathrm{calc.}}$	 &	$b_{\alpha}^{\mathrm{expt.}}$ &   &   \\[1pt]
$J^\pi$ & (MeV)  & $\alpha_{IC}$ &  (MeV)  &    (MeV)     &    (\%)   &   (\%)  &  $T_{1/2}^{\mathrm{calc.}}$   &  $T_{1/2}^{\mathrm{expt.}}$   \\[2pt] \hline
&  &  &  &  &  &  &  &  \\[-6pt]
$0^{+}$	 & 0.0000 &         &    &  $1.550 \times 10^{-15}$  &  100.000  &  100  &  294.3 ns  &  294.3(8) ns  \\
$2^{+}$	 & 0.7273 & 0.01393 &  $9.406 \times 10^{-11}$  &  $4.061 \times 10^{-14}$  &  0.043  &  0.033  &  4.8 ps  &  14.2(18) ps  \\
$4^{+}$	 & 1.1325 & 0.0543  &  $7.418 \times 10^{-12}$  &  $8.224 \times 10^{-14}$  &  1.097  & $\approx 0.5$  &  60.8 ps  & 69(10) ps \cite{TRS2021,XUNDL}\tnote{a}  \\
$6^{+}$	 & 1.3555 & 0.324   &  $4.880 \times 10^{-13}$  &  $3.273 \times 10^{-14}$  &  6.285  &  $\approx 3(1)$  &  0.88 ns  &  0.76(21) ns\tnote{b}  \\
$8^{+}$	 & 1.4764 & 3.248   &  $6.994 \times 10^{-14}$  &  $4.226 \times 10^{-15}$  &  5.698  &  $\approx 3(1)$  &  6.2 ns  &  14.6(3) ns  \\[2pt] \hline
\end{tabular}
\begin{tablenotes}
\item [a] The recent experimental work of Karayonchev {\it et al.}~\cite{KRJ2022} presents the measure $T_{1/2}^{\mathrm{expt.}}(4^{+}) = 70(5)$ ps for $^{212}$Po.
\item [b] The recent experimental works of von Tresckow {\it et al.}~\cite{TRS2021} and Karayonchev {\it et al.}~\cite{KRJ2022} present the measures $T_{1/2}^{\mathrm{expt.}}(6^{+}) = 1.15(19)$ ns and $1.14(10)$ ns for $^{212}$Po, respectively.
\end{tablenotes}
\end{threeparttable}
\end{center}
\end{table*}

Table \ref{Widths_212Po} shows the $\gamma$-decay widths ($\Gamma_{\gamma}$), $\alpha$-decay widths ($\Gamma_{\alpha}$), $\alpha$-branching ratios ($b_{\alpha}$), and half-lives ($T_{1/2}$) calculated for the states from $0^{+}$ to $8^{+}$ of the $^{212}$Po g.s.~band. The $\Gamma_{\gamma}$ values are calculated through eq.~(9) of Ref.~\cite{HMS1994}, and the $\Gamma_{\alpha}$ values are calculated by the semi-classical approximation of Ref.~\cite{GK1987}. To calculate $\Gamma_{\gamma}$, it is necessary to previously determine the $B(E2)$ transition rates for the $^{212}$Po g.s.~band; the $B(E2)$ rates calculated for $^{212}$Po are shown in Table \ref{Table_B(E2)}. Also, the value of $\Gamma_{\gamma}$ depends on the internal conversion coefficient $\alpha_{IC}$ (see values in Table \ref{Widths_212Po}), which is obtained from the internal conversion coefficient database $BrIcc$ \cite{KBT2008}. An $\alpha$ preformation factor \mbox{$P_{\alpha} = 3.902$ \%} is applied to correctly reproduce the experimental half-life of $^{212}$Po in the $0^{+}$ ground state ($T_{1/2}^{\mathrm{expt.}} = 294.3(8)$ ns); this $P_{\alpha}$ value is close to the one used by Hoyler, Mohr, and Staudt \cite{HMS1994} ($P_{\alpha} = 3.5$ \%) for the same nucleus. The $\alpha$-branching ratio is obtained by \mbox{$b_{\alpha} = \Gamma_{\alpha}/(\Gamma_{\alpha} + \Gamma_{\gamma})$}.

Analyzing the results in Table \ref{Widths_212Po}, it is noted the calculated $\alpha$-branching ratios reproduce the order of magnitude of the respective experimental data; in the case of $b_{\alpha}(6^{+})$ and $b_{\alpha}(8^{+})$, such a statement takes into account the proximity between the calculated values and the experimental uncertainty range of $\pm 1$ \%. The half-lives $T_{1/2}$ calculated for the $2^{+}$, $4^{+}$, $6^{+}$, and $8^{+}$ levels also reproduce the order of magnitude of the respective experimental values and, in the cases of $T_{1/2}(4^{+})$ and $T_{1/2}(6^{+})$, the calculated values are in complete agreement with the respective experimental data. It should be mentioned that the recent experimental work of von Tresckow {\it et al.}~\cite{TRS2021} presents the measure $T_{1/2}^{\mathrm{expt.}}(^{212}\mathrm{Po};6^{+}) = 1.15(19)$ ns, and the recent experimental work of Karayonchev {\it et al.}~\cite{KRJ2022} presents the measures $T_{1/2}^{\mathrm{expt.}}(^{212}\mathrm{Po};4^{+}) = 70(5)$ ps and $T_{1/2}^{\mathrm{expt.}}(^{212}\mathrm{Po},6^{+}) = 1.14(10)$ ns, which are still in good level of agreement with the respective calculated values.

Table \ref{Table_rms_radii} shows the rms intercluster separations ($\langle R^2 \rangle ^{1/2}$) and reduced $\alpha$-widths ($\gamma _\alpha ^2$) calculated for the levels from $0^{+}$ to $8^{+}$ of the g.s.~bands of the studied nuclei. These properties are calculated according to the procedure detailed in Subsection IV.B of our previous work in Ref.~\cite{SM2015}. For obtaining these properties, the parameter $\lambda$ is slightly modified (see $\lambda$ values in Table \ref{Table_rms_radii}) so that the radial wave function $u(r)$ of each $J^\pi$ state is calculated at the respective experimental energy, except to $^{104}$Te where the intensity parameter has the fixed value $\lambda = 0.65310$ mentioned in Table \ref{Table_parameters}; for $^{20}$Ne, $^{44}$Ti, $^{94}$Mo, and $^{212}$Po, such a procedure aims to determine the nuclear properties as $\langle R^2 \rangle ^{1/2}$, $\gamma _\alpha ^2$, $B(E2)$, etc.~in correspondence with the experimental energy levels. Analyzing the rms intercluster separations, it is noted that the antistretching effect, which is observed with other $\alpha$ + core potential shapes (e.g., \cite{SM2015,MOR1998,O1995,M2017,IMP2019}), also occurs with the potential proposed in this work; thus, the inclusion of the $c_{\mathrm{sat}} \delta(s)$ term in the effective $NN$ interaction does not change this general property. The reduced $\alpha$-widths, together with the rms intercluster separations, indicate the first levels of each g.s.~band have a stronger $\alpha$-cluster character.

\begin{table}
\caption{Rms intercluster separations ($\langle R^2 \rangle ^{1/2}$) and reduced $\alpha$-widths ($\gamma _\alpha ^2$) calculated for the $0^{+}$ to $8^{+}$ levels of the $^{20}$Ne ($G_{\mathrm{g.s.}} = 8$), $^{44}$Ti ($G_{\mathrm{g.s.}} = 12$), $^{94}$Mo ($G_{\mathrm{g.s.}} = 16$), $^{104}$Te ($G_{\mathrm{g.s.}} = 16$), and $^{212}$Po ($G_{\mathrm{g.s.}} = 24$) ground state bands. The strength parameter $\lambda$ fitted for the precise reproduction of each experimental energy is shown, except to $^{104}$Te where the strength parameter has the fixed value $\lambda = 0.65310$.}
\label{Table_rms_radii}
\begin{center}
\begin{tabular}{ccccc}
\hline
 &  &  &  &  \\[-7pt] 
  &	 &	& $\langle R^2\rangle ^{1/2}$ & $\gamma _\alpha ^2$  \\[3pt]
 Nucleus &	$J^\pi$ & $\lambda$ & (fm) & (keV) \\[2pt] \hline
 &  &  &  &  \\[-6pt]
$^{20}$Ne &   $0^{+}$  &     0.8872  &   4.018   &     39.26  \\
        &    $2^{+}$    &   0.8863   &  4.050    &    43.73  \\
        &    $4^{+}$    &   0.8872  &   4.009    &    38.65  \\
        &    $6^{+}$    &   0.8765   &  3.971    &    34.19  \\
        &    $8^{+}$    &   0.9286   &  3.500    &    6.97  \\
 &  &  &  &  \\
$^{44}$Ti  &      $0^{+}$   &    0.7945  &   4.461    &    4.32  \\
          &  $2^{+}$    &   0.7934   &  4.468    &    4.63   \\
          &  $4^{+}$    &   0.7945  &   4.427    &    3.78   \\
          &  $6^{+}$   &    0.7999  &   4.316    &    2.16  \\
          &  $8^{+}$    &   0.8034  &   4.180    &    1.02   \\
 &  &  &  &  \\
$^{94}$Mo   &    $0^{+}$   &    0.71560  &  5.158    &    0.68  \\
         &  $2^{+}$  &   0.71514  &  5.168   &   0.75  \\
         &   $4^{+}$   &   0.71560  &  5.137   &   0.60  \\
         &   $6^{+}$   &   0.71473  &  5.067   &   0.37  \\
         &   $8^{+}$  &   0.71854  &  4.949  &   0.15  \\
 &  &  &  &  \\
$^{104}$Te   &    $0^{+}$   &    0.65310  &  5.469    &    2.00  \\
         &  $2^{+}$  &   0.65310  &  5.480   &   2.19  \\
         &   $4^{+}$   &   0.65310  &  5.447   &   1.79  \\
         &   $6^{+}$   &   0.65310  &  5.351   &  1.05  \\
         &   $8^{+}$  &   0.65310  &  5.227  &   0.49  \\
 &  &  &  &  \\
$^{212}$Po   &  $0^{+}$   &   0.79005 &   6.559   &  0.63  \\
         &   $2^{+}$   &    0.78920  &  6.577   &   0.72  \\
         &   $4^{+}$   &    0.79007  &  6.556  &   0.61  \\
         &   $6^{+}$   &    0.78953 &   6.494  &   0.40  \\
         &   $8^{+}$   &    0.78940  &  6.405  &   0.21  \\[2pt] \hline
\end{tabular}
\end{center}
\end{table}

\begin{table*}
\begin{center}
\begin{threeparttable}
\caption{$B(E2)$ transition rates calculated for the levels from $0^{+}$ to $8^{+}$ of the $^{20}$Ne ($G_{\mathrm{g.s.}} = 8$), $^{44}$Ti ($G_{\mathrm{g.s.}} = 12$), $^{94}$Mo ($G_{\mathrm{g.s.}} = 16$), and $^{212}$Po ($G_{\mathrm{g.s.}} = 24$) ground state bands, \emph{without} the use of effective charges. The $B(E2)$ values of the present work are compared with results from previous studies on the $\alpha$ + core structure. Experimental data are from Ref.~\cite{ENSDF}, except where indicated.}
\label{Table_B(E2)}
\begin{tabular}{ccccccccc}
\hline
 &  &  &  &  &  &  &  &  \\[-7pt] 
 &  & \multicolumn{7}{c}{$B(E2;J\rightarrow J-2)$ (W.u.)} \\[2pt] \cline{3-9}
 &  &  &  &  &  &  &  &  \\[-8pt]
   &   &   &   &  Buck  &  Ibrahim  &  Bai \& Ren  &  Wang  &  Ni \& Ren  \\
Nucleus  &	$J^\pi$ & Expt. & This work & {\it et al.}~\cite{BMP95} & {\it et al.}~\cite{IMP2019} &
 \cite{BR2021} & {\it et al.}~\cite{WPX2013} & \cite{NR2011} \\[2pt] \hline
 &  &  &  &  &  &  &  &  \\[-8pt]
$^{20}$Ne &	$2^{+}$  &	20.3(10) &		16.646 &	14.6 &	  14.3	  &	  13.0  &  13.3  &  12.2 \\
		  &	$4^{+}$  &	22(2) 	 &		22.686 &	18.6 &	  18.5   &	  17.1  &  17.7  &  15.8  \\
		  &	$6^{+}$  &	20(3) 	 &		21.883 &	14.9 &	  15.2	  &	  15.2  &  15.2  &  13.6 \\
		  &	$8^{+}$  &	9.0(13)  &		10.834 &	7.4 &	  7.9	  &	  7.0  &  8.0  &  6.2 \\
 &  &  &  &  &  &  &  &  \\						
$^{44}$Ti &	$2^{+}$  &	22.2(+22 $-$18) \cite{ABS2017} &  11.291 &	11.0 &	10.6	&	9.9  &  10.1  &  9.3 \\
		  &	$4^{+}$  &	30(+4 $-$3) \cite{ARB2020} &	  15.423 &	14.7 &	14.1	&	13.4  &  13.7  &  12.6 \\
		  &	$6^{+}$  &	15(+3 $-$2) \cite{ARB2020} &	  14.662 &	14.1 &	13.2	&	12.7  &  13.4  &  11.8 \\
		  &	$8^{+}$  &			  &	      12.229 &		11.5 &	10.4	&	10.5  &  11.2  &  9.6 \\
 &  &  &  &  &  &  &  &  \\  
$^{94}$Mo &	$2^{+}$  &	16.0(4)	  &		8.075 	 &	   7.8  &	7.5	 &		&	 &  6.8 \\ 
		&	$4^{+}$  &	26(4)	  &		11.184	 &	  10.7  &	10.2	 &		&	& 9.3  \\
		&	$6^{+}$  &			  &		11.232	 &		     & 	10.2	 &		&	&  9.3 \\
		&	$8^{+}$  &	0.0049(8) &		9.915	 &	 	     &	9.2	 &		&	&  8.2 \\
 &  &  &  &  &  &  &  &  \\						
$^{212}$Po &	$2^{+}$  &	2.57(+38 $-$29)  &  7.521	 &			&			&	6.3  &  6.7  &  5.8\tnote{a} \\
		   &	$4^{+}$  &	9.4(13) \cite{TRS2021}\tnote{b}	 &  10.632 &			&			&	8.8   &  9.5  &  8.2\tnote{a} \\
		   &	$6^{+}$  &	13.2(+49 $-$29)\tnote{c}  &  11.035 &			&			&	9.1  &  10.0  &  8.5\tnote{a} \\
		   &	$8^{+}$  &	4.56(12)	   &  10.519 &			&			&	8.7  &  9.8  &  8.1\tnote{a} \\[2pt] \hline
\end{tabular}
\begin{tablenotes}
\item [a] $B(E2)$ rates calculated with $G = 22$.
\item [b] The recent experimental work of Karayonchev {\it et al.}~\cite{KRJ2022} presents the measure $B(E2;4^{+}_{1} \rightarrow 2^{+}_{1}) = 9.3(+7 \; -6)$ W.u.~for $^{212}$Po.
\item [c] The recent experimental works of von Tresckow {\it et al.}~\cite{TRS2021} and Karayonchev {\it et al.}~\cite{KRJ2022} present the measures $B(E2;6^{+}_{1} \rightarrow 4^{+}_{1}) = 8.7(15)$ W.u.~and $9.0(+9 \; -7)$ W.u.~for $^{212}$Po, respectively.
\end{tablenotes}
\end{threeparttable}
\end{center}
\end{table*}

Table \ref{Table_B(E2)} shows the calculated $B(E2)$ transition rates between the levels from $0^{+}$ to $8^{+}$ of the $^{20}$Ne, $^{44}$Ti, $^ {94}$Mo, and $^{212}$Po g.s.~bands. The calculated $B(E2)$ values are compared with the respective experimental data and $B(E2)$ results from previous studies \cite{BMP95,IMP2019,BR2021,WPX2013,NR2011} in which the $\alpha$ + core structure is systematically analyzed for a set of nuclei. The results of the present work provide a good general description of the experimental $B(E2)$ data, since the order of magnitude of most of the experimental data is reproduced correctly, and concerning the absolute values, there are cases of complete agreement between calculated and experimental values for the transitions: $4^{+} \rightarrow 2^{+}$ and $6^{+} \rightarrow 4^{+}$ of $^{20}$Ne; $6^{+} \rightarrow 4^{+}$ of $^{44}$Ti; $4^{+} \rightarrow 2^{+}$ and $6^{+} \rightarrow 4^{+}$ of $^{212}$Po.\footnote{The recent experimental works of von Tresckow {\it et al.}~\cite{TRS2021} and Karayonchev {\it et al.}~\cite{KRJ2022} present the measures $B(E2;6^{+}_{1} \rightarrow 4^{+}_{1}) = 8.7(15)$ W.u.~and $9.0(+9 \; -7)$ W.u.~for $^{212}$Po, respectively, which favors the theoretical predictions of Refs.~\cite{BR2021,WPX2013,NR2011}. However, the 2020 $^{212}$Po experimental data compilation \cite{AM2020,ENSDF} presents the value $B(E2;6^{+}_{1} \rightarrow 4^{+}_{1}) = 13.2(+49 \; -29)$ W.u.~shown in Table \ref{Table_B(E2)}.} Also, in the $8^{+} \rightarrow 6^{+}$ transition of $^{20}$Ne, the calculated $B(E2)$ value is very close to the uncertainty range of the respective experimental measure. Such results are rewarding, taking into account they were obtained \mbox{\emph{without}} introducing effective charges. In Table \ref{Table_B(E2)}, it is seen the $B(E2)$ values of the present work are considerably higher than those obtained in Refs.~\cite{BMP95,IMP2019,BR2021,WPX2013,NR2011} with different $\alpha$ + core potentials; consequently, the $B(E2)$ values of this work are somewhat closer to the experimental data in most transitions. Thus, the $B(E2)$ results favor the $\alpha$ + core interpretation for $^{20}$Ne, $^{44}$Ti, $^{94}$Mo, and $^{212}$Po by the methodology proposed here.

Table \ref{Table_gamma_ratios} presents the reduced $\alpha$-widths calculated for the g.s.~band of $^{212}$Po with $G_{\mathrm{g.s.}} = 22$. The alternative use of $G_{\mathrm{g.s.}} = 22$ for $^{212}$Po aims at a comparison with the $\gamma _\alpha ^2$($^{212}$Po) results obtained with $G_{\mathrm{g.s.}} = 24$ and shown in Table \ref{Table_rms_radii}. Using the $\gamma _\alpha ^2$($^{104}$Te) values, the ratios \linebreak[4] \mbox{$\gamma _\alpha ^2$($^{104}$Te)/$\gamma _\alpha ^2$($^{94}$Mo)}, \mbox{$\gamma _\alpha ^2$($^{104}$Te)/$\gamma _\alpha ^2$($^{212}$Po)} with $G_{\mathrm{g.s.}}(^{212}\mathrm{Po})$ = 22, and \mbox{$\gamma _\alpha ^2$($^{104}$Te)/$\gamma _\alpha ^2$($^{212}$Po)} with $G_{\mathrm{g.s.}}(^{212}\mathrm{Po})$ = 24 were calculated (see Table \ref{Table_gamma_ratios}). The \mbox{$\gamma _\alpha ^2$($^{104}$Te)/$\gamma _\alpha ^2$($^{94}$Mo)} values of this work are considerably higher than those shown in Ref.~\cite{SMB2019}, where the reduced $\alpha$-widths were obtained with the (1 + Gaussian)$\times$(W.S.~+ W.S.$^3$) potential; however, it should be taken into account that small variations on the length or height of the Coulomb barrier can produce strong variations in the $\gamma _\alpha ^2$ values. The use of $G_{\mathrm{g.s.}} = 22$ produces $\gamma _\alpha ^2(^{212}\mathrm{Po})$ values lower than in the case of $G_{\mathrm{g.s.}}(^{212}\mathrm{Po}) = 24$, resulting in higher \linebreak[4] \mbox{$\gamma _\alpha ^2$($^{104}$Te)/$\gamma _\alpha ^2$($^{212}$Po)} ratios for $G_{\mathrm{g.s.}}(^{212}\mathrm{Po}) = 22$. The \mbox{$\gamma _\alpha ^2$($^{104}$Te)/$\gamma _\alpha ^2$($^{212}$Po)} ratios in Table \ref{Table_gamma_ratios} are lower than those obtained in Ref.~\cite{SMB2019}; however, the calculation of Ref.~\cite{SMB2019} applies $G_{\mathrm{g.s.}}(^{212}\mathrm{Po}) = 20$, which naturally contributes to higher values of \mbox{$\gamma _\alpha ^2$($^{104}$Te)/$\gamma _\alpha ^2$($^{212}$Po)}. Even with such quantitative differences, the calculations of the present work reinforce the interpretation of Ref.~\cite{SMB2019} in which $^{104}$Te has a substantially higher degree of $\alpha$-clustering compared to $^{94}$Mo and $^{212}$Po. Such results also corroborate the analysis of Auranen {\it et al.}~\cite{ASA2018} based on experimental $^{104}$Te $\alpha$-decay data.

\begin{table*}
\caption{Reduced $\alpha$-widths calculated for the g.s.~band of $^{212}$Po ($G_{\mathrm{g.s.}} = 22$) from $J^\pi = 0^{+}$ to $8^{+}$, and ratios \mbox{$\gamma _\alpha ^2$($^{104}$Te)/$\gamma _\alpha ^2$($^{94}$Mo)}, \mbox{$\gamma _\alpha ^2$($^{104}$Te)/$\gamma _\alpha ^2$($^{212}$Po)} in the case $G_{\mathrm{g.s.}}(^{212}\mathrm{Po}) = 22$, and \mbox{$\gamma _\alpha ^2$($^{104}$Te)/$\gamma _\alpha ^2$($^{212}$Po)} in the case $G_{\mathrm{g.s.}}(^{212}\mathrm{Po}) = 24$.}
\label{Table_gamma_ratios}
\begin{center}
\begin{tabular}{ccccc}
\hline
 &  &  &  &  \\[-5pt] 
	 & $\gamma _\alpha ^2$($^{212}$Po; $G_{\mathrm{g.s.}} = 22$)  & 
\footnotesize{$\gamma_\alpha ^2(^{104}\mathrm{Te})/\gamma_\alpha ^2(^{94}\mathrm{Mo})$} & 
\footnotesize{$\gamma_\alpha ^2(^{104}\mathrm{Te})/\gamma_\alpha ^2(^{212}\mathrm{Po})$} &
\footnotesize{$\gamma_\alpha ^2(^{104}\mathrm{Te})/\gamma_\alpha ^2(^{212}\mathrm{Po})$} \\[2pt]
$J^\pi $  &  (keV)  &    & [$G_{\mathrm{g.s.}}(^{212}\mathrm{Po}) = 22$]  &  [$G_{\mathrm{g.s.}}(^{212}\mathrm{Po}) = 24$] \\[3pt] \hline
 &  &  &  &  \\[-6pt]
$0^{+}$ &  0.393  & 2.935 & 5.092  & 3.196 \\
$2^{+}$ &  0.451  & 2.906 & 4.849  & 3.050 \\
$4^{+}$ &  0.379  & 2.990 & 4.712  & 2.918 \\
$6^{+}$ &  0.241  & 2.863 & 4.369  & 2.619 \\
$8^{+}$ &  0.119  & 3.261 & 4.092  & 2.319 \\[2pt] \hline
\end{tabular}
\end{center}
\end{table*}

\section{Summary and conclusions}
\label{Sec:conclusions}

This work shows a systematic study of the $\alpha$ + core structure in $^{20}$Ne, $^{44}$Ti, $^{94}$Mo, $^{104}$Te, and $^{212}$Po nuclei using a double-folding nuclear potential with effective $NN$ interaction of \mbox{M3Y + $c_{\mathrm{sat}}\delta(s)$} type, where the term $c_{\mathrm{sat}}\delta(s)$ acts only between the $\alpha$ and core saturation regions. Using this nuclear potential plus the Coulomb term, a good description of the experimental g.s.~bands of $^{20}$Ne, $^{44}$Ti, $^{94}$Mo, and $^{212}$Po is obtained, mainly from $J^{\pi} = 0^{+}$ to $8^{+}$. With a slight variation of the $\lambda$ intensity parameter, the experimental negative parity bands of $^{20}$Ne and $^{44}$Ti are well reproduced. It is shown the additional term $c_{\mathrm{sat}}\delta(s)$ of the effective $NN$ interaction is determinant for a better reproduction of the experimental levels, and produces an effect similar to the \mbox{(1 + Gaussian)} factor of the \mbox{(1 + Gaussian)$\times$(W.S.~+ W.S.$^3$)} nuclear potential employed in previous works \cite{SM2017,SMB2019,SM2021,JQR2021}.

The proposed potential $V(r)$ is used to analyze the resonant energy levels of $^{20}$Ne and $^{212}$Po, and the $\alpha$-decay half-life of $^{104}$Te. Concerning $^{20}$Ne, the $\Gamma_{\alpha_0}$ widths calculated for the $G = 9$ negative parity band reproduce the order of magnitude of most experimental data. In the case of $^{212}$Po, the $\alpha$-branching ratios and half-lives calculated for the g.s.~band reproduce the order of magnitude of the experimental data with a preformation factor $P_{\alpha} \approx 3.9$ \%, and there is complete agreement with the experimental half-lives of the $4^{+}$ and $6^{+}$ states. With respect to $^{104}$Te, the model shows agreement with the experimental measurements of $Q_{\alpha}$ and $T_{1/2,\alpha}$ in the energy range $5.044 \; \mathrm{MeV} < E(0^{+}) < 5.3 \; \mathrm{MeV}$ using an $\alpha$ preformation factor $P_{\alpha} = 1$; this result corroborates the indication of superallowed $\alpha$-decay of $^{104}$Te in previous studies.

The $B(E2)$ transition rates calculated for the g.s.~bands of $^{20}$Ne, $^{44}$Ti, $^{94}$Mo, and $^{212}$Po present a good level of agreement with the experimental data, as the order of magnitude of most experimental $B(E2)$ rates is reproduced correctly, and there is complete agreement with the experimental data in some transitions without the use of effective charges. The $B(E2)$ results of the present work are somewhat closer to the experimental rates than other systematic calculations on the $\alpha$ + core structure in previous studies.

The calculated reduced $\alpha$-widths are used to determine the ratios \mbox{$\gamma _\alpha ^2$($^{104}$Te)/$\gamma _\alpha ^2$($^{94}$Mo)}, \mbox{$\gamma _\alpha ^2$($^{104}$Te)/$\gamma _\alpha ^2$($^{212}$Po)} with $G_{\mathrm{g.s.}}(^{212}\mathrm{Po}) = 22$, and \mbox{$\gamma _\alpha ^2$($^{104}$Te)/$\gamma _\alpha ^2$($^{212}$Po)} with $G_{\mathrm{g.s.}}(^{212}\mathrm{Po}) = 24$; the obtained values reinforce the interpretation of Souza {\it et al.}~\cite{SMB2019} in which $^{104}$Te has an $\alpha$-clustering degree substantially higher than $^{94}$Mo and $^{212}$Po, and corroborate the analysis by Auranen {\it et al.}~\cite{ASA2018} based on experimental $^{104}$Te $\alpha$-decay data.

In conclusion, the double-folding nuclear potential with M3Y + $c_{\mathrm{sat}}\delta(s)$ effective $NN$ interaction provides a good general description of energy levels, $B(E2)$ rates, $\alpha$-widths and $\alpha$-decay half-lives through a systematics applied to the set \{$^{20}$Ne, $^{44}$Ti, $^{94}$Mo, $^{104}$Te, $^{212}$Po\}. Additionally, this calculation allows reproducing the effect of the \linebreak[4] \mbox{(1 + Gaussian)} factor of the (1 + Gaussian)$\times$(W.S.~+ W.S.$^3$) nuclear potential from the viewpoint of the double-folding model. The successful use of the $c_{\mathrm{sat}}\delta(s)$ term in the effective $NN$ interaction suggests the importance of a further discussion on this phenomenological approach in terms of subnucleonic interactions.

\begin{acknowledgement}
The authors thank the HPC resources provided by Information Technology Superintendence (HPC-STI) of University of S\~{a}o Paulo.
Support from Instituto Nacional de Ci\^{e}ncia e Tecnologia -- F\'{\i}sica Nuclear e Aplica\c{c}\~{o}es (INCT-FNA) is acknowledged.
\end{acknowledgement}

%
%

\end{document}